\documentclass[twocolumn,showpacs,preprintnumbers,amsmath]{revtex4}
\usepackage{epsfig}

\begin{document}

\title{Solar Flare Intermittency and the Earth's Temperature Anomalies}
\author{Nicola Scafetta$^{1,2}$ and Bruce J. West$^{1,2,3}$}
 
\address{$^{1}$Pratt School EE Dept., Duke University,  P.O. Box 90291, 
Durham, NC 27708 }
   \address{$^{2}$ Physics Department, Duke University, Durham, NC 27708}
   \address    {$^{3}$ Mathematics Division,
Army Research Office, Research Triangle Park, NC 27709. } 
  
   \date{\today}
\date{\today}

\begin{abstract}
We argue that earth's short-term temperature anomalies and the  solar flare
intermittency are linked. The analysis is based upon the study of the
scaling of both the spreading and the entropy of the diffusion generated by the fluctuations of 
the temperature time series. The joint use of these two methods evidences the
presence of a L\'{e}vy component in the temporal persistence of the
temperature data sets that corresponds to the one that would be induced by
the solar flare intermittency. The mean monthly temperature datasets cover
the period from 1856 to 2002.
\end{abstract}


\maketitle


The historical recognition that the sun warms the earth has suggested a
direct connection between the average global temperature and solar activity.
Consequently any significant changes in solar activity should result in
equivalent changes in the earth's global temperature. The literature on the
solar influence on the earth's temperature is quite extensive \cite
{tante,Baranyi,fix}, indicating the importance of the problem and that there
are many issues that require further investigation. Herein we address the
relation between the statistics of solar flare activity and the fluctuations
in the earth's global temperature.

The dynamics of the sun's surface is turbulent, as is evidenced by changes
in solar flare activity, with 11 or 22 year solar cycles \cite{Baranyi,fix}
and strong erratic fluctuations associated with solar flare intermittency 
\cite{fix,solarflares,vulpiani,vulpiani2}. Solar irradiance changes in
accordance with the frequency of solar flares because each flare releases
more energy than the background irradiance \cite{fix}. This time variation
in the frequency of solar flares induces a similar pseudo-periodic cycle in the
earth's average temperature, as well as produces trends that move the
global temperature up or down for tens or even hundreds of years \cite
{tante,Baranyi,fix}. However, it is less evident that short-term (weekly and monthly)
changes in the global temperature are tied to solar activity whose short-time fluctuations would have the
intermittent dynamics of solar flares. Traditional measures, such as
cross-correlation functions, would not show the connection between
short-time fluctuations in the global temperature and solar flare activity,
because of the strong nonlinear hydrodynamic interactions across the earth's
surface. For example, the hydrodynamic interaction of the atmosphere over
land and water, would suppress any direct correlation between the
intermittent sun's irradiance and the earth regions' short-time response.

This letter focuses on an alternate approach to establishing the connection
between the sun's irradiance and the earth's temperature fluctuations. A
link between the two phenomena is detected through a detailed scaling
analysis of the time series for the earth's temperature and the time series
for the solar flare frequency.

Considering a solar flare as an event, the time series for the number of
solar flares has been interpreted as a waiting time distribution function
between events. The solar flare waiting time distribution function is
determined to be an inverse power-law probability density function (iplpdf) 
\cite{solarflares,vulpiani,vulpiani2}. Such dynamical stochastic processes
can be described by generalizations of random walks. A \textit{L\'{e}%
vy-flight}, for example, is such a process with a diverging second moment. A 
\textit{L\'{e}vy-walk}, on the other hand, visits the same spatial sites as
does a L\'{e}vy-flight, but each step takes a finite time and the second
moment is finite. The time necessary to complete a step in a L\'{e}vy-walk
is specified by an inverse power-law waiting time distribution function, as
first noticed by Shlesinger et al. \cite{shlesinger}. It has been determined  \cite{solarflares} that a L\'{e}vy-walk can
describe the intermittent solar flare signal. Scafetta et al. \cite
{scalingdetection,dea4} established that, in general, the presence of a L%
\'{e}vy-walk process in a given time series can be detected by the joint use
of two separate scaling techniques, the Diffusion Entropy Analysis (DEA) and
Standard Deviation Analysis (SDA). We apply the same approach to the
analysis of temperature data sets and compare them with the L\'{e}vy-walk
statistics induced by solar flare intermittency.

We study the earth's temperature anomalies both globally and locally, that is,
North-South hemispheres and Land-Ocean regions, see Fig. 1.
The technical term \emph{temperature anomalies} denotes temperature
departures from the 1961-1990 mean temperature value. These data are
recognized in the geophysical community as among the most accurate data
files for global air temperature, and global sea surface temperature (SST) 
\cite{jones}. The basic data set of global earth's temperature anomalies
(HadCRUT) is a combination of Land air temperature anomalies \cite{jones1}
(CRUTEM1) and sea-surface temperature anomalies \cite{parker} on a 5%
${{}^\circ}$ x 5${{}^\circ}$ grid-box basis. The merging of the two data
sets is discussed in Parker et al. \cite{parker} and more recently in Jones
et al. \cite{jones11}. The Land surface time series are calculated using
data from the Global Historical Climatology Network (Version 2) and
sea-surface temperature anomalies from the United Kingdom MOHSST data set
and the NCEP Optimum Interpolated SSTs (Version2). The data that we analyze
are mean monthly temperature anomalies downloaded from Climatic Research
Unit, UK, \cite{web1} (Global and North-South temperatures, 1856-2002) and
from National Climatic Data Center, USA, \cite{web2} (Land-Ocean
temperatures, 1880-2002). 

\begin{figure}[tbp]
\epsfig{file=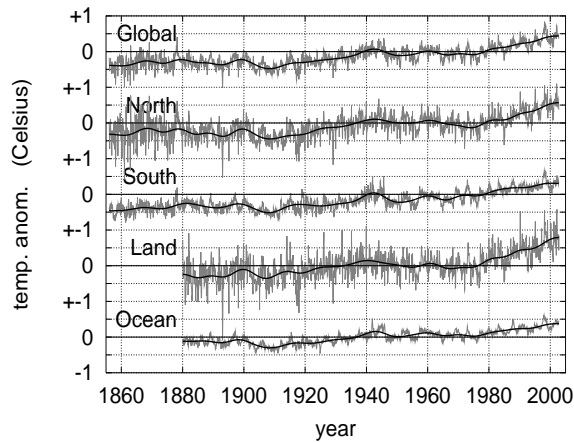,height=8cm,width=6.0cm,angle=-90}
\caption{Global and local Earth temperature anomalies in Celsius degree
(years: 1856-2002). }
\end{figure}

The analysis of the above data sets is done with both DEA and SDA
techniques. These methods are based on the prescription that the numbers in
a time series $\{\xi _{i}\}$ are the fluctuations of a diffusion trajectory;
see Refs. \cite{solarflares,scalingdetection,dea4} for details. Therefore,
we shift our attention from the time series $\{\xi _{i}\}$ to the
probability density function (pdf) $p(x,t)$ of the corresponding diffusion
process. Here $x$ denotes the variable collecting the fluctuations and is
refered to as the diffusion variable. The scaling property of $p(x,t)$ takes
the form 
\begin{equation}
p(x,t)=\frac{1}{t^{\delta }}~F\left( \frac{x}{t^{\delta }}\right) ~,
\label{scafun12}
\end{equation}
where $\delta $ is the scaling exponent. The DEA \cite{scalingdetection} is
based on the evaluation of the Shannon entropy, $S(t)$, using the pdf (\ref
{scafun12}). If the scaling condition of Eq. (\ref{scafun12}) holds true, it
is easy to prove that the entropy is 
\begin{equation}
S(t)=-\int\limits_{-\infty }^{\infty } p(x,t) \ln[p(x,t)]=A+\delta ~\ln (t)~,  
\label{scafun14}
\end{equation}
where $A$ is a constant. The SDA \cite{scalingdetection}, instead, is based
on the evaluation of the standard deviation $D(t)$ using the same pdf (\ref
{scafun12}) and yields to 
\begin{equation}
D(t)=\sqrt{\left\langle x^{2};t\right\rangle -\left\langle x;t\right\rangle
^{2}}\propto t^{H}~,  \label{varvar33}
\end{equation}
where $H$ is the Hurst exponent \cite{scalingdetection,2Mandelbrot}. 
\begin{figure}[tbp]
\epsfig{file=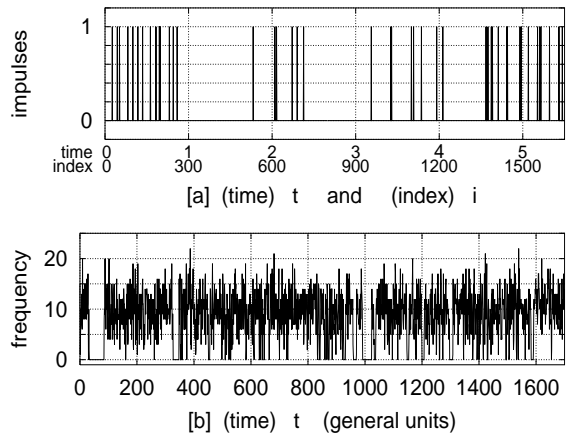,height=8cm,width=6.0cm,angle=-90}
\caption{ Fig. 2a shows a computer-generated intermittent sequences of
impulses. Fig. 2b shows the frequency of impulses for 1700 time units. }
\end{figure}

The L\'{e}vy-walk statistics is a particular form of anomalous diffusion
obtained by generalizing the Central Limit Theorem \cite{gnedenko}.
While all theoretical details can be found in Refs. \cite
{solarflares,scalingdetection,dea4} and in the enclosed references, here we
show the properties of a L\'{e}vy-walk using a model that simulates solar
flare intermittency. We generate an artificial sequence $\{\tau _{j}\}$
distributed according to an $iplpdf$, 
\begin{equation}
\psi (\tau )\propto \frac{1}{(T+\tau )^{\mu }}~,
\label{densitydistributionoftau}
\end{equation}
where $T$ is a positive constant. In the computer-generated example, we use $%
T=90$ and $\mu =2.5$, that is, we use $2<\mu <3$, so as to violate the
constraint of a finite variance on the central limit theorem that is necessary for obtaining L\'evy-walk statistics. The sequence
of waiting times $\{\tau _{j}\}$ is used to generate a time series $\{\xi _{i}\}$
given by $\xi _{i}=1$ only at the occurrence of the events, that is, for $%
i=[\sum_{j=1}^{m}\tau _{j}]$; $[a]$ is the integer part of $a$. However, in
the interval between two events, that is, for $[\sum_{j=1}^{m}\tau
_{j}]<i<[\sum_{j=1}^{m+1}\tau _{j}]$, we assume $\xi _{i}=0$. As shown in
Ref. \cite{solarflares,scalingdetection,dea4} by using results of Refs. \cite
{gnedenko,zumofenklaftershelsinger,feller} the time series $\{\xi _{i}\}$
generates a diffusion process with a scaling exponent $\delta =1/(\mu -1)$.
Moreover, this particular approach to L\'{e}vy statistics yields a diffusion
process with finite second moments, as in the L\'{e}vy-walk \cite{shlesinger}%
. Consequently, it is possible to evaluate the scaling exponent of the
standard deviation, $H$, via Eq. (\ref{varvar33}). Fig. 2a shows the first 1650 data points in the computer-generated sequence $%
\{\xi _{i}\}$, which correspond to the first 5.5 time units, of a 510,000
long data sequence $\{\xi _{i}\}$. In Fig. 2b we show the sequence of
frequencies $\{f_{t}\}$ of impulses for each time unit where we use 1 time
unit = 300 natural units of $\tau $. 

The analysis is based on the comparison of the scaling exponents $\delta $
and $H$ measured by DEA and SDA respectively. It has been shown \cite
{scalingdetection,dea4} that if a time series is characterized by Gaussian
statistics, the two scaling exponents are identical, that is, $\delta =H$.
If, instead, the process under study is characterized by a L\'{e}vy-walk,
the two scaling exponents $\delta $ and $H$ can be both related to the
exponent $\mu $ of the waiting time $iplpdf$ of the underling intermittent
process generating the L\'{e}vy statistics. In the L\'{e}vy-walk case the
scaling exponent $H$ is slightly larger than $\delta $ and the two exponents
are related to $\mu $ and one another via the following relation \cite
{scalingdetection,dea4} 
\begin{equation}
\delta =\frac{1}{3-2H}=\frac{1}{\mu -1}~.  \label{relHdelta34}
\end{equation}
We refer to (\ref{relHdelta34}) as the L\'{e}vy-walk diffusion relation. The
assessment of the property expressed by Eq. (\ref{relHdelta34}), that is, the comparison between the value of
the scaling exponents $\delta $ and $H$ measured for the earth's temperature
data sets and their relation with the scaling exponents $\delta $ and $\mu $
measured for the hard x-ray solar flares waiting time series \cite
{solarflares}, is used in this letter to verify the conjecture that the
earth's global temperature inherits the intermittent L\'{e}vy-like nature of
solar flares, within the temporal range from few weeks to few months. 

\begin{figure}[tbp]
\epsfig{file=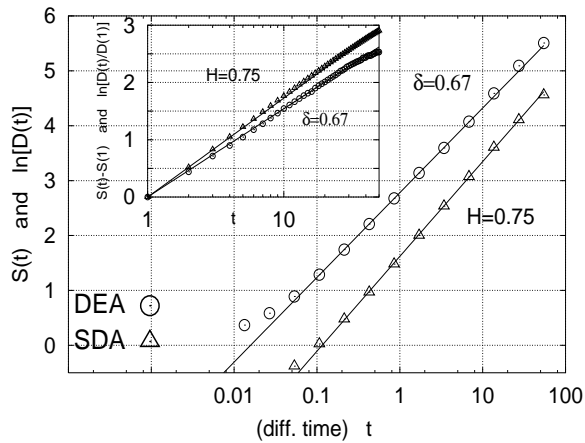,height=8cm,width=6.0cm,angle=-90}
\caption{DEA and SDA of the computer-generated L\'{e}vy-walk sequences shown
in Figs. 2a and 2b. The two straight lines show the two theoretical scaling
coefficients $\delta =0.67$ and $H=0.75$ for $\mu =2.5$.}
\end{figure}

In the computer-generated model we interpret the fluctuations in the frequency of the pulses
shown in Fig. 2b as inducing a similar memory pattern in the  monthly mean
global temperature fluctuations. Infact, a higher solar flare frequency
implies a higher monthly solar activity. The DEA and SDA of the computer-generated
data are shown in Fig. 3. These data are fit with Eqs. (\ref{scafun14}) and (%
\ref{varvar33}) yielding the two scaling coefficients $\delta =0.67$ and $%
H=0.75$. These coefficients are consistent with $\mu =2.5$ obtained by using
Eq. (\ref{relHdelta34}); the known input value for this realization of the
model. The large picture in Fig. 3 refers to the analysis of the original
510,000 data-point sequence $\{\xi _{i}\}$. The inset refers to the analysis
of 1700 data-point frequency sequence $\{f_{t}\}$ shown in Fig. 2b, a length
compatible with the length of the mean monthly temperature datasets here
analyzed. The two graphs in Fig. 3 show very good agreement with L\'{e}%
vy-walk theory, that is, with Eq. (\ref{relHdelta34}) because both time series produce similar diffusion trajectories. In particular,  the small inset in Fig.
3 shows the typical \textit{bifurcation} associated to the two scaling laws
that characterize L\'{e}vy-walk statistics and that the statistics generated by only N = 1700 frequency data
points is rich enough to obtain a satisfactory pdf with L\'evy-walk scaling properties. 

Fig. 4 shows the numerical results by applying DEA and SDA to the
temperature fluctuation data of the four regions of the earth. On the
ordinate axis we plot $\ln [D(t)/D(1)]$ and $S(t)-S(1)$; thus, all curves
start from 0. The scaling exponents $\delta $ and $H$ are obtained by
fitting the first 10 points. This analysis for the Global temperature (not shown in the figure) yields $\delta =0.90\pm 0.02$ and 
$H=0.95\pm 0.02$. The high values of the exponents imply a strong
persistence in the temperature fluctuations. Both scaling
exponents $\delta $ and $H$ are slightly larger for the South hemisphere
than for the North hemisphere and for the Ocean than for the Land. A
reasonable explanation is that the Ocean is an almost stationary system with
a very high effective heat capacity. The Land, instead, is subject to
stronger random fluctuations (as Fig. 1 shows) due to a lower effective heat
capacity and a higher morphological variability due, for example, to the
presence of deserts, forests, mountains and valleys. Finally, we note that
the standard deviation scaling exponents $H$ are larger than the diffusion
entropy scaling exponents $\delta $ and seem to fulfill the L\'{e}vy-walk
diffusion relation (\ref{relHdelta34}) within the accuracy of our
statistical analysis. 
\begin{figure}[tbp]
\epsfig{file=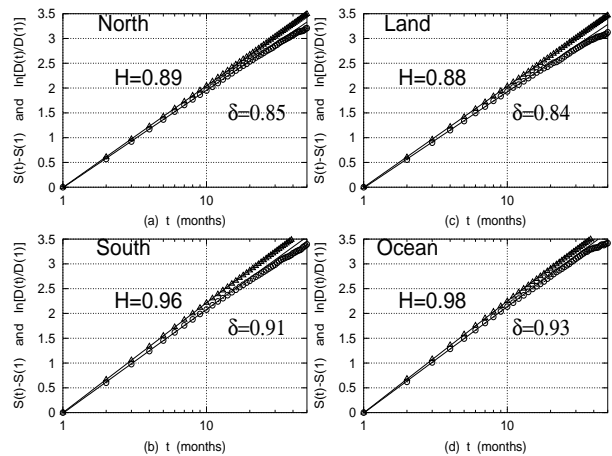,height=8cm,width=6.0cm,angle=-90}
\caption{DEA and SDA of the North-South hemisphere and Land-Ocean region
temperature anomalies. The scaling exponent $H$ and $\delta $ are reported
in the figure, the error is $\pm 0.02$. }
\end{figure}

The scaling behavior in the temperature anomalies can be related to the
intermittency of the solar flare activity through the mechanism suggested by
the model depicted in Figs. 2 and 3. In fact, the waiting time distribution of solar flares is
characterized by an $iplpdf$ of the type of Eq. (\ref
{densitydistributionoftau}) with $\mu =2.14\pm 0.05$ \cite{solarflares}. This value of $\mu $
would imply a L\'{e}vy-walk with $\delta =0.88\pm 0.02$ and $H=0.93\pm 0.02$
(\ref{relHdelta34}) for a shuffled dataset that would destroy any temporal correlation among flares and conserve only the L\'evy component; see Ref. \cite
{solarflares} about the effect of the shuffling on these data. We observe
that the difference of the above two scaling exponent values gives $H-\delta =0.05\pm 0.02,$ that
is a value compatible with the difference between $H$ and $\delta $ measured for all
the temperature data sets. In particular, $H-\delta =0.05\pm 0.02$ for the
Global temperature and for both South and Ocean regions and $H-\delta
=0.04\pm 0.02$ for North and Land regions. 
\begin{figure}[tbp]
\epsfig{file=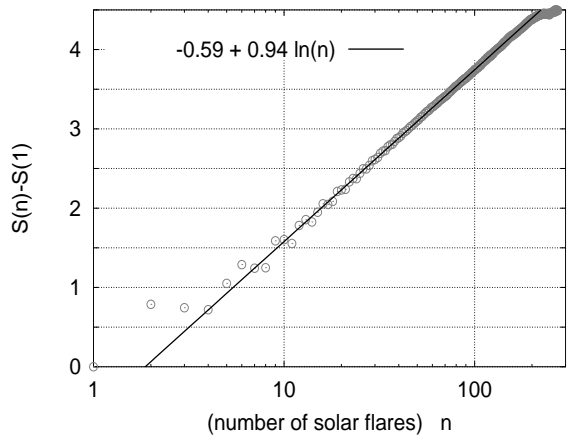,height=8cm,width=6.0cm,angle=-90}
\caption{DEA of the unshuffled waiting time of solar flares sequence during
the period 1991-2000. The value of the scaling exponent is $\delta =0.94\pm
0.02$. }
\end{figure}

To understand the actual values that we measure for $H$ and $\delta $ we
first notice that the waiting time sequence of solar flares is weakly
correlated \cite{solarflares}. Therefore, the effective value of $\delta $
for the original unshuffled sequence is slightly larger than that calculated
via the waiting time $iplpdf$ exponent $\mu $ for flares \cite{solarflares}, $%
\delta =0.94\pm 0.02$, as Fig. 5 shows. We observe that this value of $%
\delta $ is perfectly compatible with $\delta =0.93\pm 0.02$ measured for
the Ocean temperature. The Ocean, in fact, is the region of the earth that
should best mirror solar activity because it is the largest and most
homogeneous region of the earth with
a very high effective heat capacity; properties that retain correlated events in
the temperature. Also we observe that the occurrence of uncorrelated events
would slightly reduce the L\'{e}vy-like memory and increase the Gaussianity
of the data. This might explain why the scaling exponents $H$ and $\delta $ as well as their difference
slightly decreases (0.04 against 0.05) between the North and Land regions.
Finally, Fig. 5 shows that the scaling properties for the waiting time of solar
flares last at least  200 consecutive flares and the waiting time for
clusters of 200 solar flares may last for several months. The largest
waiting time between two consecutive solar flares, during the period
1991-2000, is almost 3 months \cite{solarflares}, and on average there are almost 60 solar flares per month. Therefore, the temporal
resolution of these clusters is compatible with the temporal range of
scaling of the temperature data sets that last at least 1 to almost 20
months, as Fig. 4 shows.

In conclusion, the affinity of the scaling exponents obtained through our
analysis suggest that the earth's temperature anomalies inherits a L\'{e}%
vy-walk memory component from the intermittency of solar flares. We obtain
reliable scaling properties within a short time interval, ranging from a few
weeks to a few months. Morever, the joint use of SDA and DEA has proved to be
very useful in detecting a significant L\'{e}vy component in the time series
and, in general,  it may suggest the presence of a link between a phenomenon under
study and an intermittent process characterized by a waiting time $iplpdf$.

 {\bf Acknowledgment:} N.S. thanks the ARO for support under grant DAAG5598D0002.


\end{document}